# Photonic Localization of Interface Modes at the Boundary between Metal and Fibonacci Quasi-Periodic Structure

Xiao-Ning Pang, Jian-Wen Dong*, and He-Zhou Wang**

State Key Laboratory of Optielectric Materials and Technologies, Zhongshan (Sun Yat-sen) University, Guangzhou 510275, China

We investigated on the interface modes in a heterostructure consisting of a semi-infinite metallic layer and a semi-infinite Fibonacci quasi-periodic structure. Various properties of the interface modes, such as their spatial localizations, self-similarities, and multifractal properties are studied. The interface modes decay exponentially in different ways and the highest localized mode is found to be a mode in the lower stable gap with the largest gap width. A localization index is introduced to understand the localization properties of the interface modes. We found that the localization index of the interface modes in some of the stable gaps will converge to two slightly different constants related to the parity of the Fibonacci generation. In addition, the localization-delocalization transition is also found in the interface modes of the transient gaps.

*dongjwen@mail.sysu.edu.cn, ** stswhz@mail.sysu.edu.cn

**I. INTRODUCTION**

During the past few decades, there is an increasing interest in the research of interface modes in the heterostructure which is composed of two different kinds of structures. Interface modes were first known as surface states in the research of electronic surface states in solid-state physics. Analogous to surface states in the electronic systems, optical surface states in the stratified periodic structures, known as one-dimensional (1D) photonic crystals (PCs), were previously studied [1,2]. The wave vector of surface electromagnetic (EM) waves exceeds that of light in vacuum. The excitation of this kind of surface wave needs the help of attenuated total reflection due to the boundary matching condition. The EM waves in both sides (1DPC and air) decay exponentially away from the boundary. Actually, such surface waves are interface modes at the boundary between 1DPC and air. The studies on this kind of interface modes have been extended to other photonic heterostructure with a PC and a homogenous dielectric medium [3-5].

There is another kind of interface modes that can be excited under the condition of zero in-plane wave vector. It exists in a heterostructure formed by metal and dielectric Bragg reflector. Analogous to the Tamm state in solid-state physics, this kind of interface mode is called optical Tamm state [6-8]. In fact, it has been proven that the interface mode is present when the summation of wave impedances of both reflectors in the heterostructure goes to zero, no matter the excitation is inside or outside the light cone [9-11]. From the impedance matching point of view, it is easy to understand the interface modes in two rigorous periodical PC-PC configurations [12-14] can be excited in the zero in-plane wave vector condition. However, there are few of studies on interface modes in the nonperiodic structure.

Quasicrystals are the nonperiodic structures that are constructed following a simple deterministic generation rule and have interesting optical properties [15,16]. Quasicrystals of the Fibonacci type are interesting for the wave transport and localization studies in electronic and photonic systems [17,18]. Fibonacci structure exhibits an energy spectrum that consists of a self-similar Cantor set with zero Lebesgue measure [19], and has forbidden frequency regions similar to photonic band gaps [20,21]. In contrast with the Anderson localization, the light waves in quasicrystals are critically localized which follows most likely a power-law rather than exponential decay [22,23]. However, there are few studies on interface modes in this fascinating class of structures in electronic and photonic systems [24-26]. Many localization behaviors of this kind of interface modes such as the localization-delocalization transition and self-similarities and multifractals are still unclear, which do not exist in the periodic ones. Besides, there is no study on interface modes at the boundary between a metallic layer and a Fibonacci structure.

In this paper, we will examine the existences of the interface modes in such kind of heterostructure. We observe rich and interesting localization properties of the interface modes, such as the spatial localizations, self-similarities, and multifractal properties.

The interface modes decay exponentially in different ways. In particular, a mode in the lower stable gap with the largest gap width is found to be highest spatially localized. We employ a localization index which enables us to understand the behaviors of localizations. We find that the limit of the localization index of some of interface modes in the stable gaps is related to the parity of Fibonacci generation number, while the limit of the others goes to a unique constant. We also find the localization-delocalization transition of the interface modes in the transient gaps.

This paper is organized as follows. In Sec. II, the geometry and the material parameters of the heterostructure, as well as the transfer matrix method that is employed to obtain the photonic band structure and the eigen-frequency of the interface modes, are described. In Sec. III, the numerical results and discussions are presented. Finally, the conclusions are in Sec. IV.

## II. PHYSCIAL STRUCTURE AND ANALYSIS METHOD

The structure under study is a heterostructure constructed by a semi-infinite metallic layer and a semi-infinite structure where each unit cell is composed of a Fibonacci generation, as shown in Fig. 1. In each Fibonacci unit cell, two dielectric layers A and B are stacked alternatively, according to the recursion relation, $S_{j+1} = \{S_j, S_{j-1}\}$, $S_1 = A$, $S_2 = AB$, where $j$ is the generation number of the Fibonacci unit cell. There are $F_j$ layers in $S_j$, where $F_j$ is a Fibonacci number given recursively as $F_{j+1} = F_j + F_{j-1}$, for $j \geq 1$, with $F_0 = F_1 = 1$. The layers A and B are two kinds of dielectric layers with the refractive indices of $n_A = 2.35$ and $n_B = 1.38$. Both layers are the quarter-wave plates, i.e. $n_{A(B)} d_{A(B)} = \frac{\pi}{2} \frac{\hbar c}{\hbar \omega_0}$ where $c$ is the speed of light, and $\hbar \omega_0$ is the central frequency. The metallic layer is frequency-dependent following the Drude model with a plasma frequency of $9.2 eV$. For simplify, we assume that the metal is lossless.

The dispersion relation of the Fibonacci quasi-periodic structure is derived by using transfer matrix methods [2]. Take $z$ axis as the direction normal to the surface of the heterostructure, and assume that it is homogeneous and isotropic in the $x$-$y$ plane. The EM field in the structure can be described as,

$$E_y(z) = \begin{cases} a_l e^{ik_{lz}(z-z_l)} + b_l e^{-ik_{lz}(z-z_l)} & z \geq 0 \\ C e^{\alpha z} & z < 0 \end{cases}, \quad (1)$$

$$H_x(z) = -\frac{1}{i\omega\mu} \cdot \begin{cases} ia_l k_{lz} e^{ik_{lz}(z-z_l)} - ib_l k_{lz} e^{-ik_{lz}(z-z_l)} & z \geq 0 \\ C\alpha e^{\alpha z} & z < 0 \end{cases}, \quad (2)$$

where $\alpha = \sqrt{k_y^2 - \varepsilon_M \omega^2/c^2}$ is the radiating extinction coefficient of the metal, $k_{lz} = \sqrt{\varepsilon_l \omega^2/c^2 - k_y^2}$ is the wave vector of $z$ direction in the $l$-th layer ($l$ is the layer index), $z_l$ is the position of the front surface of the $l$-th layer, and $\omega$ is the angular frequency. In this paper, we only consider the case of zero in-plane wave vector, thus we have $k_y = 0$. In the matrix presentation, by defining $a_l^+ = a_l + b_l$, $a_l^- = (a_l - b_l)/i$ [27], the EM field between the first layer and the ($F_j + 1$)-th layer in a Fibonacci unit cell can be connected by the transfer matrix $M^{(j)}$, yielding

$$\begin{pmatrix} a_{F_j+1}^+ \\ a_{F_j+1}^- \end{pmatrix} = M^{(j)} \begin{pmatrix} a_1^+ \\ a_1^- \end{pmatrix} = \begin{pmatrix} m_{11}^{(j)} & m_{12}^{(j)} \\ m_{21}^{(j)} & m_{22}^{(j)} \end{pmatrix} \begin{pmatrix} a_1^+ \\ a_1^- \end{pmatrix}, \quad (3)$$

where $M^{(j)}$ obeys the recursion relation, $M^{(j)} = M^{(j-2)}M^{(j-1)}$. The dispersion relationship (band structure) of the *j*-th Fibonacci structure can be found when $\left| \left( m_{11}^{(j)} + m_{22}^{(j)} \right) \right| = 2$. Furthermore, applying the boundary conditions at $z = 0$, it is straightforward to obtain the condition of the interface modes,

$$m_{12}^{(j)}\eta_{metal}^2 - \left( m_{11}^{(j)} - m_{22}^{(j)} \right)\eta_{metal} - m_{21}^{(j)} = 0 \qquad (4)$$

where $\eta_{metal} = \frac{\alpha}{k_{1z}} = \frac{\alpha}{k_{Az}}$. The advantage to use the coefficients $\left( a_l^+, a_l^- \right)$ is that we can find the solutions of Eq. (4) in real space, instead of complex space if we use the coefficients $\left( a_l, b_l \right)$. One can extract all the information about the interface modes and obtain the distribution of the eigen-frequency in different Fibonacci structures.

As the matrix $M^{(j)}$ is a periodic function, the whole spectrum of the Fibonacci structure is repeated. We can obtain $F_j$ eigenvalues and $F_j$ eigenvectors for each generation within a given frequency interval. The eigenvector for a particular eigenvalue gives the spatial distribution of the interface mode. As the generation increases, the numbers of eigenvectors, as well as the information contained in the eigenvectors, are overwhelming. It would be useful to consider some characteristic numbers that tell us something useful about the spatial properties of the modes. The localization index $\gamma$ is introduced to evaluate the localization properties of each mode [28], which is defined as,

$$\gamma = \frac{\sum_{i=1}^{F_j} I_i^2}{\left( \sum_{i=1}^{F_j} I_i \right)^2} \qquad (5)$$

where $I_i$ is the average intensity of each layer. The value of $\gamma$ varies significantly between localized and extended modes. The larger localization index is, the more localized is. If the intensity profile is uniformly distributed, the localization index reaches its minimum value of $1/F_j$. On the other hand, if the intensity profile is localized in a single layer, the localization index reaches its maximum value of 1.

### III. RESULTS AND DISCUSSIONS

Figure 2(a) shows the band structure and the eigen-frequency of the interface modes as a function of the generation numbers from $j=2$ to $j=8$. Indeed, with our choice of the central frequency ($\hbar\omega_0 = 1.0\,eV$), the whole band structure is repeated periodically every $2.0\,eV$, and the band structure is symmetry about $\hbar\omega_0$. There are two kinds of photonic band gaps in the Fibonacci structure: one is the transient gap which appears every three generation (e.g. near 1.0 eV when $j=2,5,8$); another is the stable gap which appears every generation and locates on both sides of the corresponding transient gap (e.g. near 0.8 and 1.15 eV when $j \geq 4$). It is seen that the width of the stable gaps remains almost constant for all generations, whereas the width of the transient gap becomes narrow with the increasing of the generation number. This is consistent with the previous findings [26]. Similar evolution patterns of the band structures will repeat in the lower and higher frequency regions due to the self-similarity of Fibonacci structures. The self-similarity properties are also found in the band structures of higher generation structures [see more in Figs. 2(b), 2(c), and 2(d)]. On the other hand, Fig. 2(a) shows that there is an interface mode (marked by red point) in each photonic band gap and the distribution of the eigen-frequencies is similar to that of the band structures, except that it is no longer symmetry about $\hbar\omega_0$. Different symbols are used to clarify the modes in different gaps. In order to describe in convenience, we denote the modes in the transient gaps and the stable gaps as the stable interface modes (SIMs) and the transient interface modes (TIMs). We note that there are two kinds of SIMs due to the different asymptotic behaviors of their localization indices, which will be discussed below. We also note that the completely tunneling states appear at the frequency of $2.0\ eV$ in every generation because the matrix $M^{(j)}$ becomes an identity matrix in this special case.

Fig. 3 shows the intensity profiles of a TIM and two SIMs in log-linear scale. The resonant frequencies of these three modes are 0.9985, 0.8011, and 1.1535 eV in the 8th Fibonacci structure. We can find that all of the intensity profiles exponentially decay away from the interface of the heterostructure. The SIM with a resonant frequency of 0.8011 eV is the most spatial localized than the others. This is because such resonant frequency is nearest to the center of the stable gap with the largest gap width. The EM waves suffer strong instructive interferences and decay significantly. For the other two modes, since they locate near the band edge [see in Fig. 2(a)], their EM behaviors are affected by the extended states of the Fibonacci structures. As a result, the EM waves can penetrate into the Fibonacci structure more easily. Insets in Fig. 3 are the magnified intensity profiles. The self-similar of intensity profiles can be confirmed by comparing the inset and the whole profile.

We have also calculated the localization index of the interface modes as a function of the number of layers so as to study the localization behaviors. The results are shown in Fig. 4, and their resonant frequencies are around 1.0, 0.8, 1.15 eV. We find that the variation of the localization index is highly dependent on the modes. For the TIM, its localization index decreases as the Fibonacci generation increases. When the Fibonacci generation increases, we already see that the width of the transient gap becomes more and more narrow, and goes to vanish in the high generation limit. In other words, the TIM experiences the transition from localization to delocalization as the Fibonacci size increases.

However, such a transition behavior will not occur in the SIM. The reason is that the widths of the stable gaps in the Fibonacci structures, as well as the resonant frequencies of the SIMs, are almost the same in every generation (see Fig. 2). In fact, there are two kinds of SIMs with different localization characteristics. One of the SIMs is the well-defined localized modes due to their intensity profile invariance. It can be demonstrated, from Fig. 4(b), that the localization index first drops a little at $j=5$ and then remains almost a constant regardless of the generation of the Fibonacci structure. We denote them as SIM-I. In addition, we find that the localization index of the SIM at around $0.8 eV$ is the largest, showing that it is the most localized. However, another kind of SIMs (so-called SIM-II) is also localized in space, but the localization index is much lower than those of the SIM-I [see Figs. 4(b) and 4(c)]. What's more, the localization index goes to two constants with a little differences related to the parity of Fibonacci generation number, and the modes in the odd generations are more localized than those in the even generations [see Fig. 4(c) for detail]. This parity dependence will affect a lot to the spatial localization properties. When we change $\omega_p$ from 9.2 to 4.0 eV, we find that the resonant frequencies of the SIM-II in the odd Fibonacci structure will blue-shift obviously, while those in even Fibonacci structure almost keep invariant. But the TIM and the SIM-I are changed regardless of the parity.

We also do the multifractal analysis to describe the statistical properties of the interface modes. The method is followed by the previous literatures [26,29]. The multifractal spectrum for the TIM, as shown in Fig. 5(a), varies in a finite region of singularity strength $[\alpha_{min}, \alpha_{max}]$. The range of singularity strength $\alpha_{max} - \alpha_{min}$ can be used to reflect the randomness of the intensity profile. This is another evidence to show that the TIMs possess multifractal properties that are unique characteristics of self-similar modes, and the mode is something in between localized and extended. Contrary to the TIM, the multifractal spectrum for the SIM-I [Fig. 5(b)] and the SIM-II (results not shown here) exhibits the characteristic of localized waves, namely $\alpha_{min}$ goes to zero and $\alpha_{max}$ increases rapidly as the generation number increases.

## IV. CONCLUSIONS

In summary, the existence and localization of the interface modes in the heterostructure have been investigated. The heterostructure is composed of a semi-infinite metal and a semi-infinite Fibonacci quasi-periodic structure. Rich localization characteristics of the interface modes, such as their spatial localizations, self-similarities, and multifractal properties, are well studied. The interface modes are classified into three kinds of interface modes according to their localization properties. The localization index is employed to analyze their localization behaviors. With the help of such index, we found that, in particular, the mode in the lower stable gap with the largest gap width is highest spatially localized. The interface modes in the transient gaps will transit from localization to delocalization with the increasing of Fibonacci generation number; while the modes in the stable gaps are localized due to their localization index going to constants in higher order generations. In addition, some of the indices of the stable interface modes have two limits with respect to the even/odd Fibonacci generation. The expected self-similarity of the interface modes is also found in the mode patterns and the frequency distributions of the modes.

**ACKNOWLEDGEMENT**

We would like to thank Dr. C. P. Yin and T. B. Wang for their valuable discussions. This work is supported by the National Natural Science Foundation of China (10804131, 10874250, 10674183) and the Fundamental Research Funds for the Central Universities (2009300003161450).

**Figure Captions**

Fig. 1. (Color online) The schematic diagram of the 1D heterostructure consisting of a semi-infinite Drude-model metal and a semi-infinite structure where each unit cell is composed of the 5th generation Fibonacci generation. Both layers A and B is the quarter-wave plates.

Fig. 2. (Color online) (a) The band structures (blue line segments) and the eigen-frequencies (red symbols) with different generation numbers from $j=2$ to $j=8$. Full circles (●) represents the modes in the transient gap; while full lower triangles (▼) and full upper triangles (▲) represent the modes in the stable gaps with the localization index of one and two limits, respectively. Others are marked by open circles (○). (b)-(d) Same as (a) but for higher generations in the frequency around 1.65, 1.0, and 0.35 eV in order to show the self-similarity.

Fig. 3. (Color online) The intensity profiles of (a) the TIM, (b) the SIM-I, and (c) the SIM-II in the heterostructure composed of the 8th generation Fibonacci structure. Their resonant frequencies are 0.9985, 0.8011, and 1.1535 eV, respectively. The red lines guide the eyes to indicate the exponential decay. Insets show part of the intensity profiles for clarity.

Fig. 4. The relationship between the localization index $\gamma$ and the number of layers $F_j$ for (a) the TIM, (b) the SIM-I, and (c) the SIM-II. Their resonant frequencies are at around 1.0, 0.8, and 1.15 eV. The solid curves guide the eyes.

Fig. 5. (Color online) The multifractal spectra for (a) the TIM at around 1.65 eV and (b) the SIM-I at around 0.8 eV.

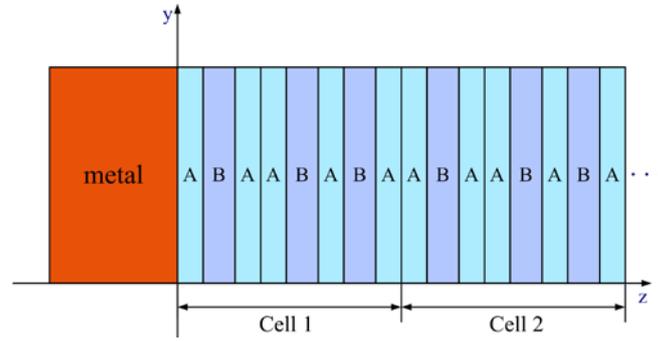

**Fig. 1. (Color online) The schematic diagram of the 1D heterostructure consisting of a semi-infinite Drude-model metal and a semi-infinite structure where each unit cell is composed of the 5th generation Fibonacci generation. Both layers A and B is the quarter-wave plates.**

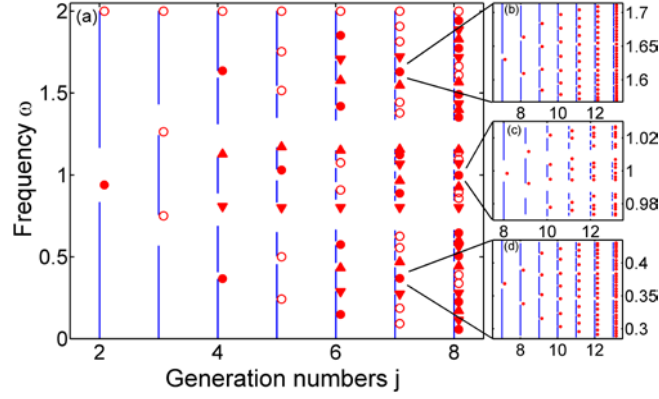

**Fig. 2. (Color online) (a) The band structures (blue line segments) and the eigen-frequencies (red symbols) with different generation numbers from $j=2$ to $j=8$. Full circles (●) represents the modes in the transient gap; while full lower triangles (▼) and full upper triangles (▲) represent the modes in the stable gaps with the localization index of one and two limits, respectively. Others are marked by open circles (○). (b)-(d) Same as (a) but for higher generations in the frequency around 1.65, 1.0, and 0.35 eV in order to show the self-similarity.**

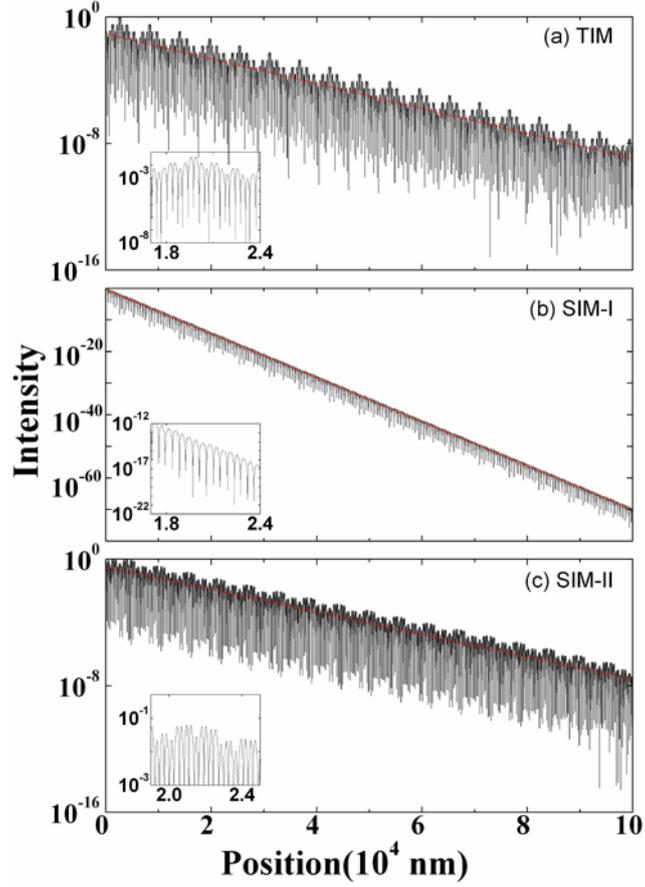

**Fig. 3. (Color online) The intensity profiles of (a) the TIM, (b) the SIM-I, and (c) the SIM-II in the heterostructure composed of the 8th generation Fibonacci structure. Their resonant frequencies are 0.9985, 0.8011, and 1.1535 eV, respectively. The red lines guide the eyes to indicate the exponential decay. Insets show part of the intensity profiles for clarity.**

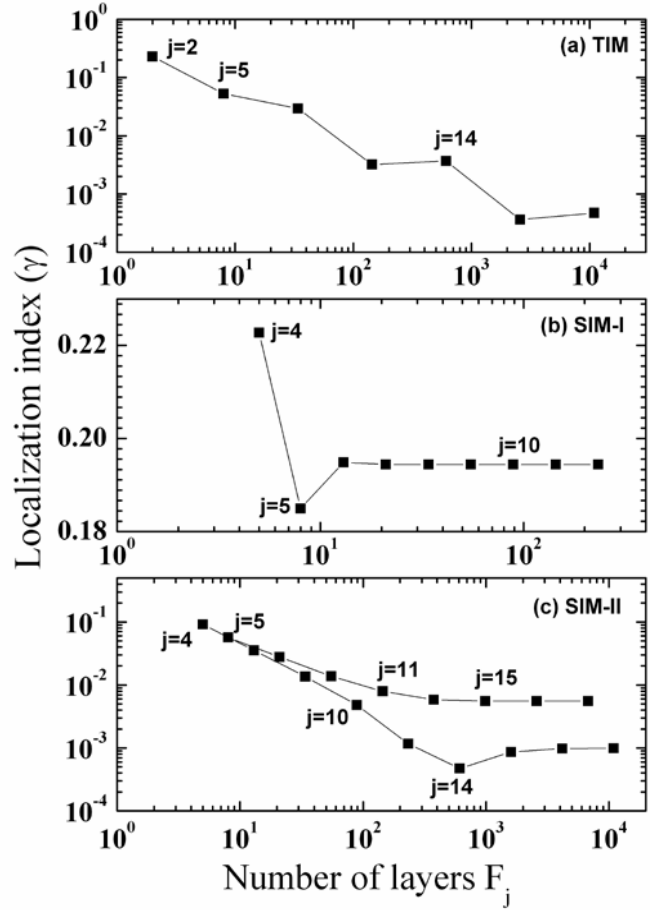

Fig. 4. The relationship between the localization index $\gamma$ and the number of layers $F_j$ for (a) the TIM, (b) the SIM-I, and (c) the SIM-II. Their resonant frequencies are at around 1.0, 0.8, and 1.15 eV. The solid curves guide the eyes.

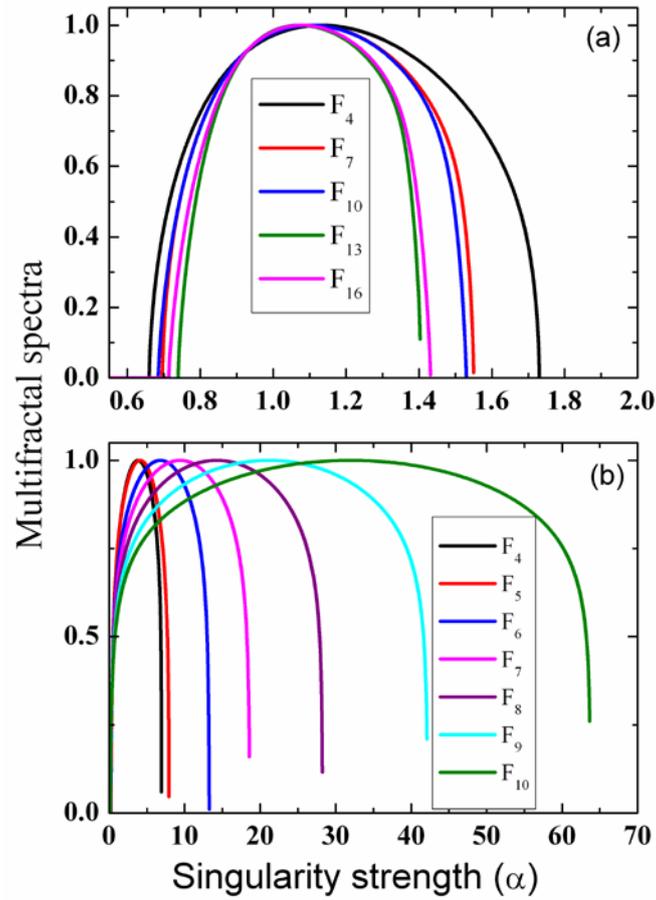

**Fig. 5. (Color online) The multifractal spectra for (a) the TIM at around 1.65 eV and (b) the SIM-I at around 0.8 eV.**